\documentclass[12pt]{article}
\usepackage{epsfig}
\input epsf.sty
\newcommand{\la}[1]{\label{#1}}
 
\newcommand{\be}{\begin{equation}}
\newcommand{\ee}{\end{equation}}
\newcommand{\ba}{\begin{eqnarray}}
\newcommand{\ea}{\end{eqnarray}}
\newcommand{\bi}{\begin{itemize}}
\newcommand{\ei}{\end{itemize}}

\newcommand{\nr}[1]{(\ref{#1})}
\newcommand{\tr}{{\rm Tr\,}}

\newcommand{\msbar}{\overline{\mbox{\rm MS}}}
\newcommand{\lambdamsbar}{\Lambda_{\overline{\rm MS}}}

\newcommand{\bmu}{\bar{\mu}}

\newcommand{\RR}{{\rm I\kern -.2em  R}} 
\newcommand{\eq}{Eq.~}

\newcommand{\fig}{Fig.~}

\def\lsi{\raise0.3ex\hbox{$<$\kern-0.75em\raise-1.1ex\hbox{$\sim$}}}
\def\gsi{\raise0.3ex\hbox{$>$\kern-0.75em\raise-1.1ex\hbox{$\sim$}}}
\newcommand{\lsim}{\mathop{\lsi}}

\def\none               {\multicolumn{2}{c|}{---}}

\makeatletter \@addtoreset{equation}{section} \makeatother

\makeatletter
\renewcommand\section{\@startsection {section}{1}{\z@}%
                                   {-5.5ex \@plus -1ex \@minus -.2ex}
                                   {2.3ex \@plus.2ex}%
                                   {\normalfont\large\bfseries}}
\renewcommand\subsection{\@startsection{subsection}{2}{\z@}%
                                     {-3.25ex\@plus -1ex \@minus -.2ex}%
                                     {1.5ex \@plus .2ex}%
                                     {\normalfont\normalsize\bfseries}}
\renewcommand\thesection {\@arabic\c@section}
\renewcommand\thesubsection   {\thesection.\@arabic\c@subsection}
\renewcommand{\@seccntformat}[1]{%
\csname the#1\endcsname.\hspace{1.0em}}
\makeatother

\begin{document}
 
\begin{titlepage}
\begin{flushright}
Edinburgh 2000/18\\
CERN-TH/2000-300\\
hep-lat/0010008\\
\end{flushright}
\begin{centering}
\vfill
 
{\bf TESTING IMAGINARY VS. REAL CHEMICAL \\
POTENTIAL IN FINITE-TEMPERATURE QCD}

\vspace{0.8cm}
 
A. Hart$^{\rm a,}$\footnote{Current address: 
DAMTP, University of Cambridge, Cambridge CB3 0WA, UK},
M. Laine$^{\rm b,c}$, 
O. Philipsen$^{\rm b,}$\footnote{Current address: 
Center for Theoretical Physics, MIT, Cambridge, MA 02139, USA}

\vspace{0.3cm}
{\em $^{\rm a}$%
Dept.\ of Physics and Astronomy, Univ.\ of Edinburgh,\\
Edinburgh EH9 3JZ, Scotland, UK\\}
\vspace{0.3cm}
{\em $^{\rm b}$%
Theory Division, CERN, CH-1211 Geneva 23,
Switzerland\\}
\vspace{0.3cm}
{\em $^{\rm c}$%
Dept.\ of Physics, P.O.Box 9, FIN-00014 Univ.\ of Helsinki,
Finland\\}

\vspace*{0.7cm}
 
\end{centering}
 
\noindent
One suggestion for determining the properties of QCD at finite 
temperatures and densities is to carry out lattice simulations with 
an imaginary chemical potential whereby no sign problem arises, and 
to convert the results to real physical observables only afterwards. 
We test the practical feasibility of such an approach for a particular 
class of physical observables, spatial correlation lengths in the 
quark-gluon plasma phase. Simulations with imaginary chemical potential 
followed by analytic continuation are compared with simulations with 
real chemical potential, which are possible by using a dimensionally 
reduced effective action for hot QCD (in practice we consider QCD 
with two massless quark flavours). We find that for imaginary 
chemical potential the system undergoes a phase transition at 
$|\mu/T|\approx \pi/3$, and thus observables are analytic only 
in a limited range. However, utilising this range, relevant 
information can be obtained for the real chemical potential case. 
\vfill
\noindent
 

\vspace*{1cm}
 
\noindent
Edinburgh 2000/18\\
CERN-TH/2000-300\\
February 2001 

\vfill

\end{titlepage}
 

\section{Introduction}

Given the applications to cosmology and heavy 
ion collision experiments, it is important to determine the 
properties of QCD at finite temperatures and 
baryon densities. For instance, one would like to know 
the locations of any phase transitions, and the
properties of the quark-gluon plasma phase such as its 
free energy density, or pressure, as well as the spatial
and temporal correlation lengths felt by various 
types of excitations in the system. 

Because the theory is strongly coupled, the only 
practical first principles method available 
for addressing these questions is lattice simulations. 
While there has been steady improvement in the accuracy of 
results at vanishing baryon density~\cite{lat99}, the case
of a non-vanishing density is still largely open, despite 
much work~\cite{lat99}--\cite{jsk}. 
Indeed, introducing a non-vanishing
density, or chemical potential, is difficult because it leads
to a measure which is not positive definite (this is the so called 
sign problem), whereby standard Monte Carlo techniques fail. 

In this paper we focus on one of the suggestions for how
a finite density system could eventually be addressed with
practical lattice simulations. The idea is to first inspect an 
imaginary chemical potential, whereby the sign problem temporarily
disappears, and then relate this to the case of a real chemical potential. 
Let us denote by $\mu$ the chemical potential
for quark number $Q$, and by  $\mu_B$  the chemical potential
for baryon number $B=Q/3$: then $\mu = \mu_B/3$. By 
$\mu_R,\mu_I\in \RR$ we denote the
real and imaginary parts of $\mu$:
\be
\mu = \mu_R + i \mu_I.
\ee
It is easy to see, by going to momentum space, that physical 
observables are periodic in $\mu_I$ with the period $2\pi T$.

There are then two types of suggestions for how an imaginary 
$\mu = i \mu_I$ could be utilised 
to obtain information on a system with a real chemical
potential, $\mu=\mu_R$. The first idea is
related directly to the equation of state, 
and employs the canonical partition
function at fixed quark number~(\cite{rw,emr,dagotto,hasenfratz} 
and references therein):
\be
Z(T,Q) =  
\frac{1}{2\pi T}\int_{-\pi T}^{+\pi T} d \mu_I\, Z(T,\mu = i\mu_I)\,
e^{- i \mu_I Q/T}, \la{int} 
\ee
where the grand canonical partition function,
\be
Z(T,\mu) =  \tr e^{-(\hat H - \mu \hat Q)/T}, 
\ee
has been evaluated with an imaginary chemical potential. 
Here $\hat H$ and $\hat Q$ denote the Hamiltonian and 
quark number operators, respectively, while $Q$ is a number. 
With an imaginary chemical potential, $Z(T,\mu = i \mu_I)$
or rather the ratio $Z(T,\mu = i \mu_I)/Z(T,0)$, can be 
determined using standard lattice techniques. What remains 
is to perform the integral in \eq\nr{int}. Of course, this gets
more and more difficult in the thermodynamic limit $Q\to\infty$, 
because oscillations reappear in the Fourier transform.
In addition, a Legendre transform would be needed to go from 
$Z(T,Q)$ to a system in an ensemble 
with a real chemical potential, $Z(T,\mu = \mu_R)$.

The second idea~(\cite{mlo} and references therein) 
is that, away from possible
phase transition lines, the partition function and expectation values 
for various observables should be analytic 
in their arguments, in particular in $\mu/T$. 
Thus, we may attempt a general power series ansatz for the functional 
behaviour in $\mu/T$, 
determine a finite number 
of coefficients with an imaginary chemical potential, and finally
analytically continue to real values. 

In this letter we study the latter suggestion within the framework of
a dimensionally reduced effective theory. As we shall review 
in the next section, at temperatures sufficiently 
above the phase transition,
the thermodynamics of QCD can be represented, with good
practical accuracy, by a simple three-dimensional (3d)
purely bosonic theory. This can also be done
with a chemical potential, both real and imaginary~\cite{mu}.
We then use this theory to measure 
the longest static correlation lengths in the system for both cases.
We find that, for small $|\mu/T|$, the observables are well described by
a truncated power series with coefficients determined by fits.
We then inspect how well the analytically continued series describes the 
real data. In principle the free
energy density could be addressed with similar
effective theory methods~\cite{a0cond}, but
this requires a number of high-order perturbative 
computations which are not available at the moment for $\mu\neq 0$.

\section{Effective theory}

\subsection{Action}

The effective theory emerging from hot QCD by dimensional 
reduction~\cite{dr}--\cite{ad},\cite{mu}, 
is the SU(3)+adjoint Higgs model with the action
\be
        S = \int d^{3}x \left\{ \frac{1}{2} \tr F_{ij}^2
        +\tr [D_{i},A_0]^2 +m_3^{2} \tr A_0^2 +i \gamma_3 \tr A_0^3
        +\lambda_3(\tr A_0^2)^{2} \right\} ,  \label{action}
\ee
where $F_{ij}=\partial_{i}A_{j}-\partial_{j}A_{i}
 +ig_3[A_i,A_j]$,
$D_i = \partial_i + ig_3 A_i$, 
$F_{ij},A_i$, and $A_0$ are all traceless $3\times 3$
Hermitian matrices ($A_0=A_0^{a}T_{a}$, etc), and $g_3^2$ and $\lambda_3$
are the gauge and scalar coupling constants with mass dimension one, 
respectively.
The physical properties of the effective theory are determined by
the three dimensionless ratios 
\be
x=\frac{\lambda_3}{g_3^2}, \quad y=\frac{m_3^2(\bmu_3=g_3^2)}{g_3^4},
\quad z = \frac{\gamma_3}{g_3^3},
\la{xy}
\ee
where $\bmu_3$ is the $\msbar$ dimensional regularization scale in
3d. For vanishing chemical potential $\gamma_3 = 0$ and no term cubic
in $A_0$ appears. These ratios are via dimensional reduction functions
of the temperature $T/\lambdamsbar$ and the chemical potential
$\mu/\lambdamsbar$, as well as of the number $N_f$ of massless quark
flavours; for the case $\mu=0$, we refer to~\cite{ad}.  The inclusion
of quark masses is also possible in principle, but in the numerical
part of this work we assume $N_f=2$ massless dynamical flavours, the
other flavours being approximated as infinitely heavy.

The mass parameter $m_3^2$, represented by $y$ in \eq\nr{xy}, 
turns out to be positive~\cite{ad}. This guarantees that the 3d
theory tends to live in its symmetric phase, $A_0 \sim 0$, 
at least on the mean field level. We will return to this issue
presently.

Compared with the case $\mu = 0$, 
the dominant changes in the action
due to a small chemical potential are now~\cite{mu}
\be
z\! :\, 
0 \to  \frac{\mu}{T} \frac{N_f}{3\pi^2}; \quad
y\! :\, 
y \to y\biggl(1+\Bigl( \frac{\mu}{\pi T}\Bigr)^2 \frac{3 N_f}{2 N_c + N_f}
\biggr), \la{lo-formulas} \la{params}
\ee
where $N_c = 3$.
Thus, one new operator is generated in the effective action, 
and one of the parameters which already existed, gets modified.

For real chemical potential, $\mu=\mu_R$, the effective action 
is thus complex, whereas for imaginary chemical potential, $\mu= i \mu_I$, 
it is real.

\subsection{Ranges of validity}

There are several requirements for the effective description 
in \eq\nr{action} to be reliable. 
They are all related to a sufficiently
``weak coupling'', or effective expansion parameter,
for a given $T/\lambdamsbar, \mu/\lambdamsbar$, $N_f$. 
Let us briefly reiterate them here. 

First, the perturbative expansions for the effective 
parameters in \eq\nr{xy} have to be well convergent. 
Inspecting the actual series up to next-to-leading order, it appears
that this requirement is surprisingly well met even at temperatures
not much above the critical one~\cite{ad}.

Second, the higher dimensional 
operators arising in the reduction step which are
not included in the effective action
in~\eq\nr{action}, should only give small corrections. This condition is met
if the dynamical mass scales described by the effective theory 
are smaller than the ones  $\sim 2\pi T$ that have been integrated out. 
In pure Yang-Mills theory, there is evidence that this can be sufficiently 
satisfied at temperatures as low as 
$T\sim 2\lambdamsbar$~\cite{rold},\cite{ad}--\cite{rnew},\cite{mu}.
However, when fermions are included and a {\em real} 
chemical potential is switched on, 
some of the mass scales increase (see below), and the effective 
description will become less accurate.

Third, the effective 3d theory represents the 4d theory reliably
only when it lies in its symmetric phase~\cite{ad,su3adj}
(i.e.,\ $A_0 \sim 0$).
Indeed, for $N_f=0$ QCD has a so called Z($N$)-symmetry~\cite{gpy,ka1}, 
and this symmetry is not fully reproduced by the effective
theory. The 4d Z($N$) symmetry is however spontaneously broken
for $N_f = 0$ in the deconfined phase, and even explicitly broken 
for $N_f > 0$. 
In this case {\em broken} Z($N$) means that the Polyakov line
is approximately unity, corresponding to $A_0\sim 0$ and 
hence to a {\em symmetric} phase 
in terms of the gauge potential.
Consequently the requirement to be 
in the symmetric phase of the 3d theory
is easier to control for $N_f > 0$. 
Furthermore, the situation gets even better for real $\mu \neq 0$.
In the effective theory 
this can be seen, for instance, from the fact that the 
mass parameter $y$ in \eq\nr{params} grows, which makes it more
difficult to depart from $A_0 \sim 0$. 
Another stabilising factor is 
that the unphysical minima correspond to non-zero 
expectation values for $\tr A_0^3$~\cite{su3adj}, and the imaginary 
term $\sim i \tr A_0^3$ in the action in \eq\nr{action} disfavours 
such minima according to the standard argument~\cite{vw}. 

On the other hand, an {\em imaginary} chemical 
potential $\mu = i \mu_I$, favours those Z($N$)  
broken minima where the Polyakov line has a non-trivial 
phase, and correspondingly the gauge 
potential is non-zero, $A_0 \neq 0$. 
Utilising the perturbative effective potential~\cite{kaps},
we find that the lowest such minimum becomes 
degenerate with the symmetric one
$A_0\sim 0$ already at $\mu_I/T = \pi/3$, 
and increasing $\mu_I$ further it eventually becomes lower than our 
minimum. Thus there is a (first order) ``phase transition''~\cite{rw}. 
In the effective theory, this phase transition is triggered
by the decrease of the 
mass parameter $y$ in \eq\nr{params}. 
Moreover, in this case the term $\sim \tr A_0^3$
in the action in \eq\nr{action} favours the effective theory remnant 
of one of the minima with $A_0 \neq 0$, i.e.,\ 
a non-trivial phase of the Polyakov line.

This phase transition limits the applicability of the effective theory with 
imaginary chemical potential, 
since only the symmetric phase is a faithful 
representation of the 4d theory~\cite{ad,su3adj}. 
(For one suggestion on how perhaps to circumvent this problem
at least for $N_f=0$, 
which we shall however not dwell on here, see~\cite{bka}.) 
In a way, this problem is related to the fact that as one 
approaches $\mu_I/T  =\pi$, fermions start to obey
Bose-Einstein statistics and become ``light''
infrared sensitive degrees of freedom~(see also~\cite{kaps}), 
whereby it is no longer legitimate to integrate them out.

In summary, the effective theory roughly loses its accuracy
with a real chemical potential 
once even the longest correlation length is shorter than
$\sim 1/(2\pi T)$, and with an imaginary chemical 
potential once $|\mu/T|$ exceeds unity.
Fortunately, this range of validity contains the parameters that are 
phenomenologically most relevant. 
Indeed, heavy ion collision experiments at 
and above AGS and SPS energies can be estimated to correspond to 
$\mu_B/T \lsim 4.0$~\cite{exp}, 
or a quark chemical potential $\mu/T \lsim 1.3$.

\subsection{Observables and their parametric behaviour}

As we have mentioned, 
the physical observables which we shall study are
spatial correlation lengths: we 
consider operators living in the ($x_1,x_2$)-plane, and 
measure the correlation lengths in the $x_3$-direction.  

In the presence of $\mu \neq 0$, 
there are only two different quantum number
channels to be considered, distinguished by the two-dimensional parity $P$
in the transverse plane. The lowest dimensional gauge invariant
operators in the scalar ($J=0$) channels are:
\ba
J^{P}= 0^{+}: & & \tr A_0^2, \tr F_{12}^2,
\tr A_0^3, \tr A_0 F_{12}^2, ... 
\nonumber \\
J^{P}= 0^{-}: & & \tr F_{12}^3, \tr A_0^2F_{12}, \tr A_0 F_{12}, ...
\la{contop} 
\ea 
The corresponding 4d operators can be found in~\cite{ay}.
We shall measure whole cross correlation matrices between all (smeared) 
operators in these
channels, but mostly focus on their lowest eigenstates, 
corresponding to the longest
correlation lengths in the 4d finite temperature system.
We denote the ``energies'' of these eigenstates, viz.~inverses of 
correlation lengths, by~$m$. We also examine the overlap of 
operators of different field contents onto the eigenstates. 

Since a change $\mu\to -\mu$ can be compensated for by a field
redefinition $A_0 \to -A_0$ in \eq\nr{action}, all physical
observables must be even under this operation. In the original 4d
theory the same statement follows from compensating $\mu\to -\mu$ by a
C (or CP) operation. Moreover, since there are no
massless modes at $\mu = 0$, we expect the masses to be analytic in $\mu$
away from phase transitions.
For small values of $\mu/T$, 
the inverse correlation lengths may thus be written
as
\be \label{ansatz}
\frac{m}{T}  =  c_0 
+ c_1 \biggl( \frac{\mu}{\pi T} \biggr)^2  
+ c_2 \biggl( \frac{\mu}{\pi T} \biggr)^4
+ {\cal O}\left( \left( \frac{\mu}{\pi T} \right)^6 \right)\;.
\ee
We have chosen to include $\pi T$ in the denominators, because
the chemical potential appears 
with this structure in the effective parameters,
cf.\ \eq\nr{params}. Of course, the radii of
convergence of such expansions are not known a priori.

Here we first check to what extent 
a truncated series of the type in \eq\nr{ansatz} 
can accurately describe the data. In the range where this is possible, 
we determine the $\{ c_i \}$ with $\mu=i
\mu_I$, and check if the analytically continued result reproduces the 
independent measurements carried out with $\mu = \mu_R$. 

\section{Simulations}

\subsection{Simulation methods}

\begin{table}[tb]
\begin{center}
\begin{tabular}{|r@{.}l|r@{.}l|*{2}{r@{.}l|r@{.}l|}}
\hline
\hline
\multicolumn{2}{|c|}{} &
\multicolumn{2}{|c|}{} &
\multicolumn{4}{|c|}{real $\mu$} &
\multicolumn{4}{|c|}{imaginary $\mu$} \\
\cline{5-12}
\multicolumn{2}{|c|}{$\frac{|\mu|}{T}$} &
\multicolumn{2}{|c|}{$\frac{|\mu|^2}{(\pi T)^2}$} &
\multicolumn{2}{c|}{$y$} &
\multicolumn{2}{c|}{$z$} & 
\multicolumn{2}{c|}{$y$} &
\multicolumn{2}{c|}{$z$} \\
\hline
 0&50     & 0&0253  & 0&49218 & 0&0338     &
 0&47382 & 0&$0338i$    \\
 0&75    & 0&0570  & \none & \none & 
 0&46235 & 0&$0507i$    \\
 1&00     & 0&1013  & 0&51970 & 0&0675     & 
 0&44630 & 0&$0675i$    \\
 1&25     & 0&1583  & 0&54035 & 0&0844     & 
 0&42565 & 0&$0844i$    \\
 1&50     & 0&2280  & 0&56558 & 0&1013     & 
 0&40042 & 0&$1013i$    \\
 1&75     & 0&3103  & 0&59540 & 0&1182     & 
 \none & \none \\
 2&00     & 0&4053  & 0&62981 & 0&1351     & 
 \none & \none \\
 3&00     & 0&9119  & 0&81333 & 0&2026     & 
 \none & \none \\
 3&75     & 1&4248  & 0&99914 & 0&2533     & 
 \none & \none \\
 4&00     & 1&6211  & 1&07026 & 0&2702     & 
 \none & \none \\
\hline 
\hline
\end{tabular}
\caption{ \label{tab_params_mu}
{\em The parameters used for
$\mu\neq 0$ (cf.\ \eq\nr{xy}). 
All correspond to $T=2\lambdamsbar, N_f=2$.
In addition, $x=0.0919$, $g_3^2=2.92T$, $\beta=21$, volume $=30^3$,
where $\beta$ determines the lattice spacing
(for the detailed relations employed here, see~\cite{mu}).}}
\end{center}
\end{table}

We simulate the theory at several $\mu/T$.
The values chosen, together with the corresponding continuum parameters,
are listed in Table~\ref{tab_params_mu}. Discretization and 
lattice--continuum relations~\cite{lr} are implemented as in~\cite{mu}.
As discussed there, finite volume and lattice spacing 
effects are expected to be smaller or at most of the same order 
as the statistical errors for the parameter values we employ.
Compared with~\cite{mu}, we have increased the 
statistics and included many new values of $\mu/T$, 
in order to carry out more precise fits.

For real $\mu = \mu_R$, the action in \eq\nr{action} with parameters
as in \eq\nr{params} is complex, which precludes direct Monte Carlo
simulations. We must thus carry out simulations using a reweighting
technique, which has been explained in detail in~\cite{mu}.
There it was found that 
physically realistic lattice volumes may be simulated for chemical potentials
up to $\mu_R/T \lsim 4$.
For imaginary $\mu = i \mu_I$, 
the action in \eq\nr{action} with
parameters as in \eq\nr{params} is real, and correspondingly we
simulate the full action using a Metropolis update.

\subsection{Results}

As a first result, let us note that,
as has been the case in several related 
theories~\cite{Philipsen:1996af,hp,mu},
we again observe a dynamical decoupling of operators, such that
operators involving scalars ($\tr A_0^2, \tr A_0F_{12}^2$ etc.)
and purely gluonic operators ($\tr F_{12}^2$ etc.) 
have a mutual overlap consistent with
zero. 
The correlation
matrix thus assumes an approximately block diagonal form.  
We find that the gluonic states remain extremely insensitive to 
$\mu/T$, and agree well with the masses found in  $d=3$ pure gauge
theory~\cite{Teper:1998te}.
This situation is illustrated in~\fig\ref{fig:level}.

\begin{figure}[t]

\centerline{\epsfxsize=6cm\hspace*{0cm}\epsfbox{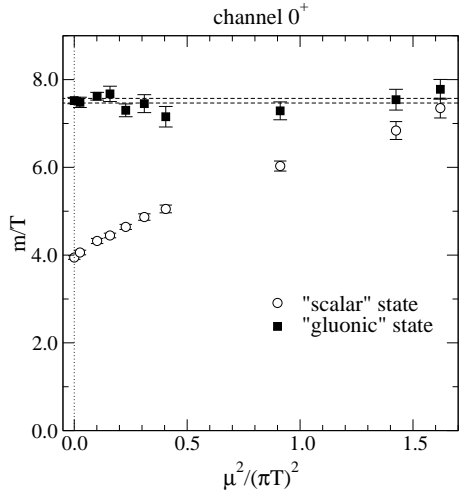}}

\caption[a]{{\em Inverse correlation lengths in the channel
$0^+$, for real $\mu/T$. ``Scalar'' states ($\tr A_0^2$ etc) 
do depend on $\mu/T$, while ``gluonic'' states ($\tr F_{12}^2$ etc)
are practically independent of it. For comparison, 
the horizontal band indicates
the 3d pure glue result for $\tr F_{12}^2$~\cite{Teper:1998te}, 
converted to our units via $g_3^2 = 2.92 T$.}}
\la{fig:level}
\end{figure}

\begin{figure}[t]

\centerline{\epsfxsize=6cm\hspace*{0cm}\epsfbox{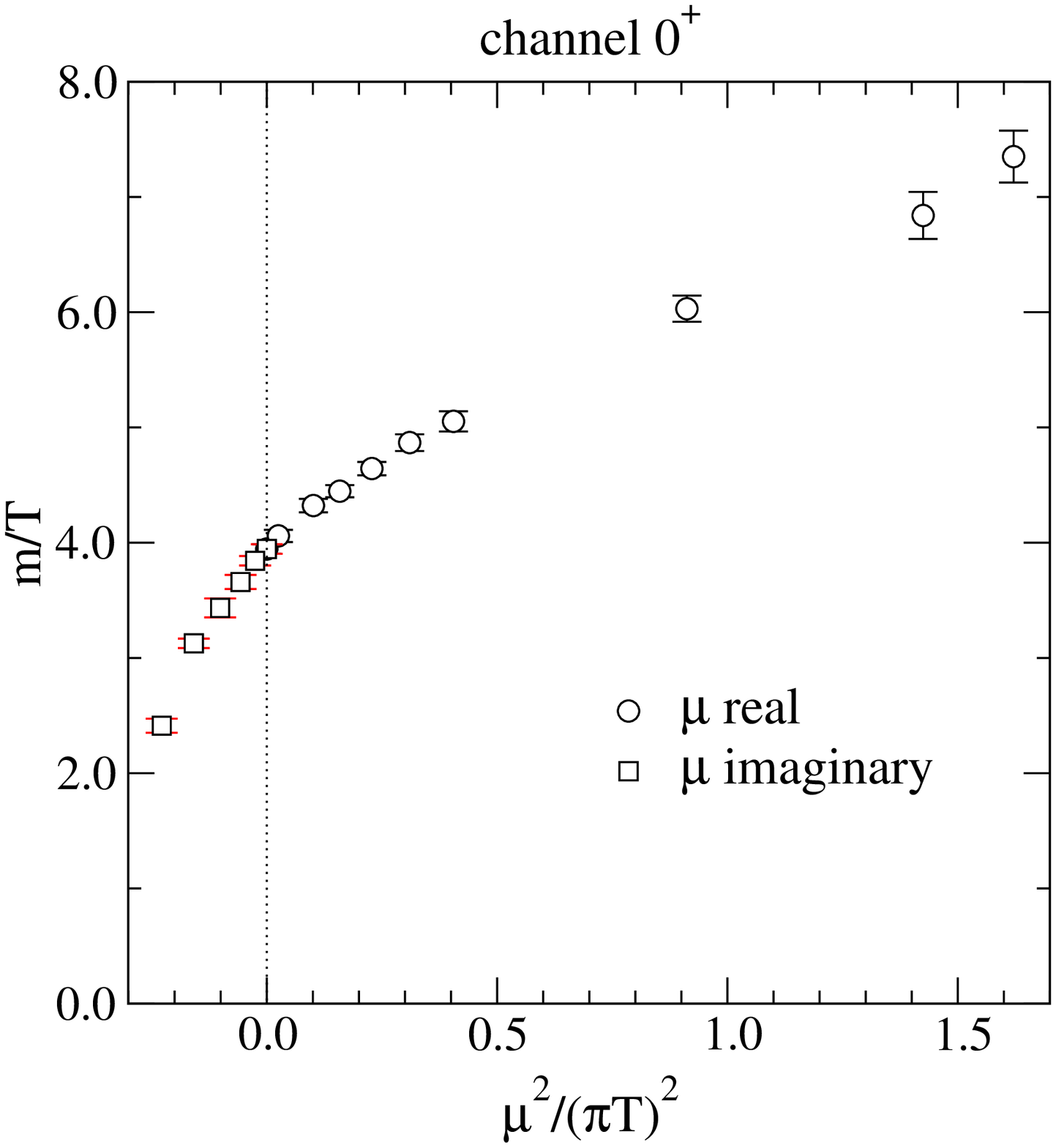}%
\epsfxsize=6cm\hspace*{1cm}\epsfbox{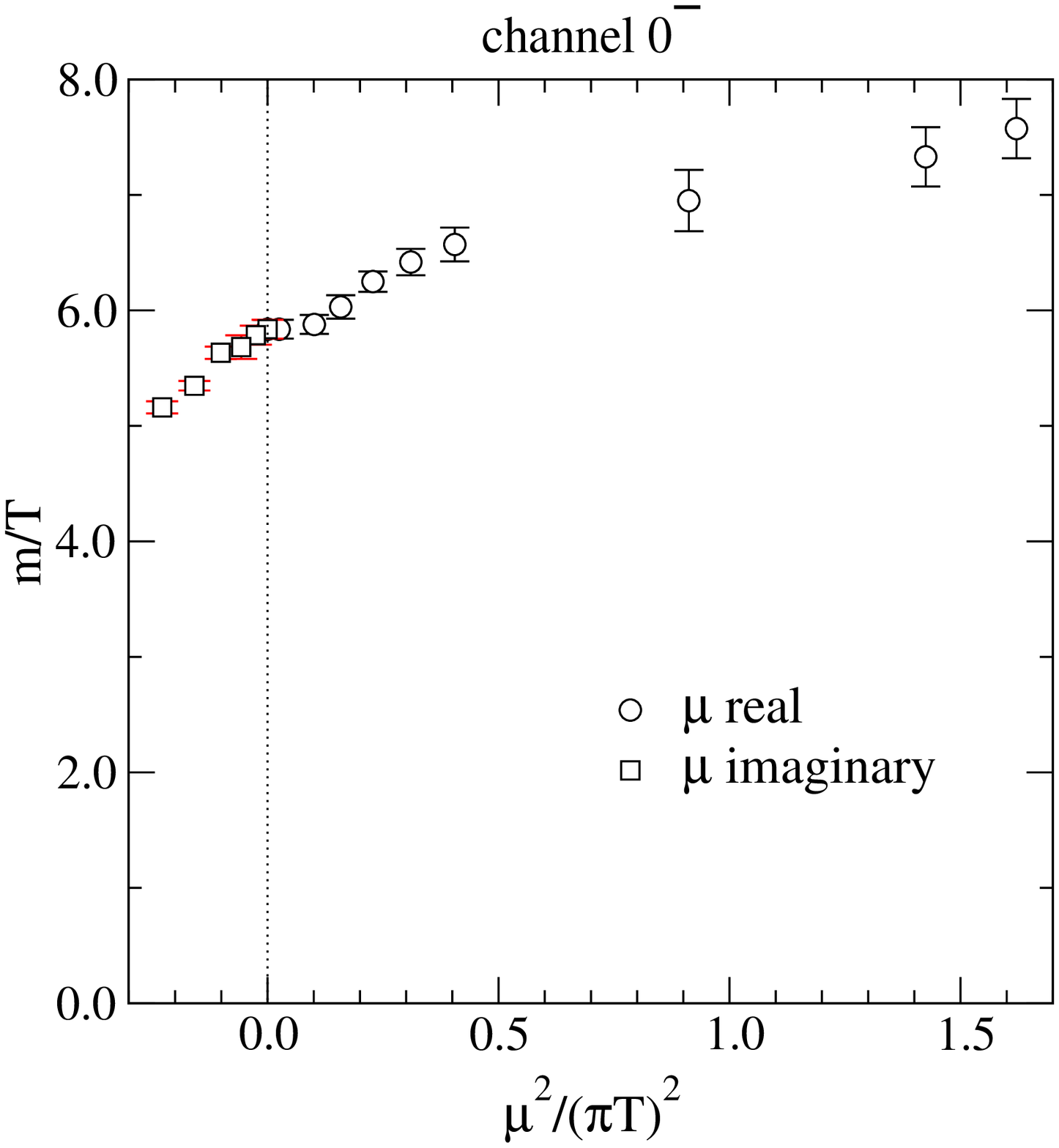}}

\caption[a]{{\em Left: Inverses of longest correlation lengths in the 
channel $0^+$, for real and imaginary $\mu$. Right: the same for 
$0^-$.}} 
\la{fig:masses}
\end{figure}

The scalar states, on the other hand, show a marked dependence
on $\mu/T$, with
their masses increasing for real $\mu$
and decreasing for imaginary
values. For both small real and small imaginary $\mu/T$, the ground
state in each channel is scalar in nature, and we plot these states 
in \fig\ref{fig:masses}.

Because of the different qualitative behaviours of 3d gluonic and 
scalar states, we may expect to observe a
change in the nature of the ground state excitation at some $\mu_R$. 
Indeed, \fig\ref{fig:level} suggests a level crossing at $\mu/T\sim 4.0$. 
This would
mean that the longest correlation length in the thermal system does
not get arbitrarily short with increasing density, but rather stays at
a constant level. Note that the value of $m/T$ at this crossing is already
so large that the effective theory 
may be inaccurate quantitatively, and in fact in the full 4d theory
the flattening off could take place much earlier.
However, the qualitative effect should be the same.

\begin{table}[tb]
\begin{center}
\begin{tabular}{|*{5}{r@{.}l|}l|}
\hline
\hline
\multicolumn{2}{|c|}{$\mu_{R}^{\rm max}/T$} & 
\multicolumn{2}{c|}{$c^R_0$} & 
\multicolumn{2}{c|}{$c^R_1$} & 
\multicolumn{2}{c|}{$c^R_2$} & 
\multicolumn{2}{c|}{$\chi^2/{\rm dof}$} & 
 $Q$ \\
\hline
 1&50 & 3&952 (37) & 3&89 (99) & $-3$&92 (449) & 0&175 & 0.840 \\
 2&00 & 3&956 (35) & 3&54 (52) & $-2$&06 (144) & 0&145 & 0.965 \\
 3&00 & 3&965 (32) & 3&22 (27) & $-1$&06 (33) & 0&216 & 0.956 \\
 4&00 & 3&983 (30) & 2&94 (20) & $-0$&61 (16) & 0&607 & 0.751 \\
\hline
\hline
\multicolumn{2}{|c|}{$\mu_{I}^{\rm max}/T$} & 
\multicolumn{2}{c|}{$c^I_0$} & 
\multicolumn{2}{c|}{$c^I_1$} & 
\multicolumn{2}{c|}{$c^I_2$} & 
\multicolumn{2}{c|}{$\chi^2/{\rm dof}$} & 
 $Q$ \\
\hline
 1&25 & 3&952 (38) & $4$&73 (157) & $-3$&07 (933) & 0&090 & 0.914 \\
 1&50 & 3&925 (35) & $2$&64 (96) & $-16$&89 (443) & 1&004 & 0.390 \\
\hline
\hline
\end{tabular}
\caption{ \label{tab_re_0p_gs} \label{tab_im_0p_gs}
  {\em Fitting the 
  lowest masses 
  in the channel $0^+$ from $\mu = 0$ up to $\mu = \mu^{\rm max}$.
  The numbers in parentheses indicate the error of the last digit shown,
  the coefficients refer to \eq\nr{ansatz}, the sub and 
  superscripts ${\scriptstyle{R,I}}$ denote real or imaginary $\mu$,  
  and $Q$ is the quality of the fit.}}
\end{center}
\end{table}

\begin{table}[tb]
\begin{center}
\begin{tabular}{|*{5}{r@{.}l|}l|}
\hline
\hline
\multicolumn{2}{|c|}{$\mu_{R}^{\rm max}/T$} & 
\multicolumn{2}{c|}{$c^R_0$} & 
\multicolumn{2}{c|}{$c^R_1$} & 
\multicolumn{2}{c|}{$c^R_2$} & 
\multicolumn{2}{c|}{$\chi^2/{\rm dof}$} & 
 $Q$ \\
\hline
 1&50 & 5&839 (69) & $-0$&54 (167) & 10&33 (722) & 0&029 & 0.971 \\
 2&00 & 5&804 (63) & 1&22 (91) & 2&06 (246) & 0&429 & 0.788 \\
 3&00 & 5&770 (57) & 2&18 (47) & $-0$&90 (65) & 0&655 & 0.658 \\
 4&00 & 5&782 (54) & 2&01 (35) & $-0$&60 (23) & 0&546 & 0.800 \\
\hline
\hline
\multicolumn{2}{|c|}{$\mu_{I}^{\rm max}/T$} & 
\multicolumn{2}{c|}{$c^I_0$} & 
\multicolumn{2}{c|}{$c^I_1$} & 
\multicolumn{2}{c|}{$c^I_2$} & 
\multicolumn{2}{c|}{$\chi^2/{\rm dof}$} & 
 $Q$ \\
\hline
 1&25 & 5&818 (71) & $0$&36 (195) & $-16$&36 (1087) & 0&298 & 0.742 \\
 1&50 & 5&857 (65) & $2$&57 (116) & $-2$&53 (465) & 0&858 & 0.462 \\
\hline
\hline
\end{tabular}
\caption{ \label{tab_re_0m_gs} \label{tab_im_0m_gs}
  {\em Fitting the 
   lowest masses in the channel $0^-$.
   The notation is as in Table~\ref{tab_re_0p_gs}.}}
\end{center}
\end{table}

Next, let us discuss the applicability of the power series 
ansatz in \eq\nr{ansatz}. 
To this end we perform fits over a range $|\mu|=0...\mu^{\rm
  max}$ to the inverses of the longest correlation lengths, 
both for real and imaginary $\mu$. For imaginary $\mu$, we
can follow the ``analytically continued'' metastable branch as
long as tunnelling into an unphysical minimum
does not become a problem, which in practice
means $\mu_I/T\lsim 1.5$.

The results are shown for the $0^+$ channel in Table~\ref{tab_re_0p_gs}, 
and for the $0^-$ channel in Table~\ref{tab_re_0m_gs}. Examining these fits
we see that in all cases we have good fits, as demonstrated by the low
$\chi^2/{\rm dof}$ and good $Q$ values. In the case of real $\mu$ we find
stable and well constrained values for the coefficients
as we increase the size of the fitting range.
For imaginary $\mu$, due to the breakdown of the effective theory at large 
values of $\mu_I/T$, we
have fewer significant data points, 
and consequently the coefficient of the quartic term
is much less constrained. 

As our main result, we can now state that we observe 
good evidence for analytic continuation in the first non--trivial
term, with $c^R_1$ consistent with $c^I_1$ in the $0^+$
channel and similarly for the $0^-$ states.
Unfortunately, the data is not accurate enough to make 
a similar statement for $c^R_2, c^I_2$. 
Extremely precise measurements would be needed,
because the range in $|\mu/T|$ available to 
imaginary chemical potential simulations is very limited. On the
other hand, from the phenomenological point of view the first
non-trivial coefficient is sufficient, since the series expansion
turns out to be in powers of $\mu^2/(\pi T)^2$, which is small 
in the most important practical applications. 
Thus, for phenomenological purposes, 
it does not seem necessary to invest an extra amount of
effort on a more precise determination of the masses in the 
imaginary $\mu$ case. 

\section{Conclusions}

In this letter, we have studied the question as to what extent
imaginary chemical potential simulations could be useful for determining
the properties of the quark gluon plasma phase at high temperatures
and finite densities. The physical observables
we have measured are static bosonic correlation lengths, 
but the pattern should be very similar for 
the free energy density, as long as $T > T_c$.

The method we have used
is based on a dimensionally reduced effective field
theory. This way we can address both a system with a real and an
imaginary chemical potential, as long as their absolute values are
relatively small compared with the temperature.
For larger absolute values of
$\mu/T = i \mu_I/T$, there is a (first order) 
phase transition, and the effective description breaks down. 

Despite the fact that
we are only working in the quark-gluon plasma phase, we find
an interesting structure in the longest correlation length, which
decreases first but becomes constant beyond some real
value of $\mu/T$, which we estimate to be $\lsim 4.0$.

Furthermore, 
in the region where the effective theory is applicable, we find that direct
analytic continuation does seem to provide a working tool for 
determining correlation lengths. For phenomenological applications, only the 
first two coefficients in the power series are needed, since we find 
the expansion parameter to be $\lsim \mu^2/(\pi T)^2$, which
is small in heavy ion collision experiments.  This is good, since
determining more coefficients with imaginary chemical potential 
would require very precise simulations. 

We are thus encouraged to believe that in 4d simulations
analytic continuation of imaginary chemical potential results would
give physically relevant results
if a good ansatz for the $\mu$-dependence is available, 
and would allow to go closer to $T_c$ determining, e.g., 
the free energy density and the spatial correlation lengths there.
Furthermore, it appears that at high enough temperatures
it is even sufficient to determine only the coefficients of the 
first terms depending on $\mu$, which amounts simply to 
susceptibility measurements at $\mu=0$ 
(see, e.g., \cite{susfree,susmass} and references therein).

\section*{Acknowledgements}

M.L. thanks Chris Korthals Altes for a useful discussion. 
This work was partly supported by the TMR network {\em Finite
  Temperature Phase Transitions in Particle Physics}, EU contract no.\ 
FMRX-CT97-0122. The work of A.H. was supported in part by UK PPARC
grant PPA/G/0/1998/00621.

\end{document}